\documentclass[journal]{IEEEtran}
\usepackage[utf8]{inputenc}
\usepackage{xcolor}
\usepackage{acronym}
\usepackage{cite}
\usepackage{amsmath,amssymb}
\usepackage{graphicx}
\usepackage{subfigure}
\usepackage{algorithm,algcompatible}
\usepackage{amsthm}
\usepackage{etoolbox}
\usepackage[justification=centering, font=small]{caption}

\algdef{SE}[SUBALG]{Indent}{EndIndent}{}{\algorithmicend\ }
\algtext*{Indent}
\algtext*{EndIndent}

\begin{document}

%\title{Beam Steering for Indoor Visible Light Multiple-Access Communications}

\title{Slow Beam Steering for Indoor Multi-User Visible Light Communications}

\author{\IEEEauthorblockN{Yusuf Said Ero\u{g}lu$^\dag$, Chethan Kumar Anjinappa$^\dag$, \.{I}smail G\"{u}ven\c{c}$^\dag$, and Nezih Pala$^\ddag$}\\
\IEEEauthorblockA{$^\dag$Department of Electrical and Computer Engineering, North Carolina State University, Raleigh, NC\\
$^\ddag$Department of Electrical and Computer Engineering, Florida International University, Miami, FL\\
Email: \{yeroglu, canjina, iguvenc\}@ncsu.edu},~npala@fiu.edu
\thanks{This work is supported in part by NSF CNS award 1422062.}
}

\maketitle
\pagenumbering{gobble}
 
\begin{abstract}
Visible light communications (VLC) is an emerging technology that enables broadband data rates using the visible spectrum. VLC beam steering has been studied in the literature  to track mobile users and to improve coverage. However, in some scenarios, it may be needed to track and serve multiple users using a single beam, which has not been rigorously studied in the existing works to our best knowledge. In this paper, considering slow beam steering where beam directions are assumed to be fixed within a transmission frame, we find the optimum steering angles to simultaneously serve multiple users  within the frame duration. This is achieved by solving a non-convex optimization problem using grid based search and majorization-minimization (MM) procedure. Additionally, we consider multiple steerable beams case with larger number of users in the network, and propose an algorithm to cluster users and serve each cluster with a separate beam. The simulation results show that clustering users can provide higher rates compared to serving each user with a separate beam, and two user clusters maximizes the sum rate in a crowded room setting.
%serving multiple users with a single steerable beam provides higher data rates compared to non-steerable beam, although serving a single user is the most efficient method in terms of the sum rates. 
\end{abstract}


\vspace{-3mm}
\section{Introduction}
Visible light communications (VLC) technology uses light sources such as LEDs for both illumination and wireless data transfer. In this technology, light-emitting diodes (LEDs) act as an antenna and transmit data to users through modulating light intensity. Due to high frequency of the modulation, the changes in the signal are not perceivable to human eye. Depending on the LED or lens type, VLC light beams can be highly directional\cite{Eroglu_JSAC, Nakagawa}. Such directional LEDs can be preferred for providing higher signal strength at longer distances, decreasing interference in other directions, or providing accurate angle of arrival information for localization purposes. 

VLC networks can provide highly accurate localization information \cite{Alphan_JLT}, and this location information can be used to steer the light beam towards user location by manipulating the orientation of the light source to further enhance the communications performance. It has been shown in the literature that using a \textit{steerable directional beam} maximizes both the overall signal strength and the coverage area \cite{Rahaim2017}. In\cite{eroglu_VLCS}, tracking users by steering LEDs is shown to provide much higher signal to interference plus noise ratio (SINR) in the VLC cell borders, which provides smoother handovers between adjacent VLC access points (APs). However, these studies assume that each user is tracked with a dedicated LED. When the number of users are lower than or equal to the number of steerable beams the steering is relatively simple, because each user can be assigned a single beam that tracks the user. However, it is highly likely that in some cases number of users are higher than the number of steerable beams. In such cases, how to steer the LEDs and distribute time allocation to users is an open problem which has not been addressed in the literature.

In this paper, we investigate the optimal beam steering parameters, especially for the case where the number of users are higher than the number of steerable beams. The optimization parameters are the steering angles, the directivity index of the LED, and the time allocation of each user. First, we define the optimization problem for a single LED and more than one user. We propose a near-optimal solution for the non-convex problem. Second, we evaluate the case where there are more than one steerable components. As a solution for beam steering and multiple access in this scenario, we propose a $k$-means clustering based user grouping algorithm. In particular, we cluster the users and assign a single beam to each cluster. Our results show that, clustering an average of two users per beam maximizes the sum rate in a 8~m~$\times$~8~m room setting with 10 users.

%The remainder of the paper is as follows. In Section~\ref{SystemModel}, we present beam steering mechanisms that are proposed in the literature for VLC and introduce the multiple access beam steering problem. In Section~\ref{PropSol}, we discuss the solution to introduced problem and extend the solution to multiple steerable beams case. In Section~\ref{SimRes} we present the simulation results, and finally in Section~\ref{Conc} we conclude the paper.

\section{System Model} \label{SystemModel}
In this section, we investigate the methods proposed for VLC beam steering with multiple access, and introduce the slow beam steering problem.

\subsection{VLC Beam Steering Methods}
\begin{figure}[tb]
	\centering
	\includegraphics[width = 3.2 in]{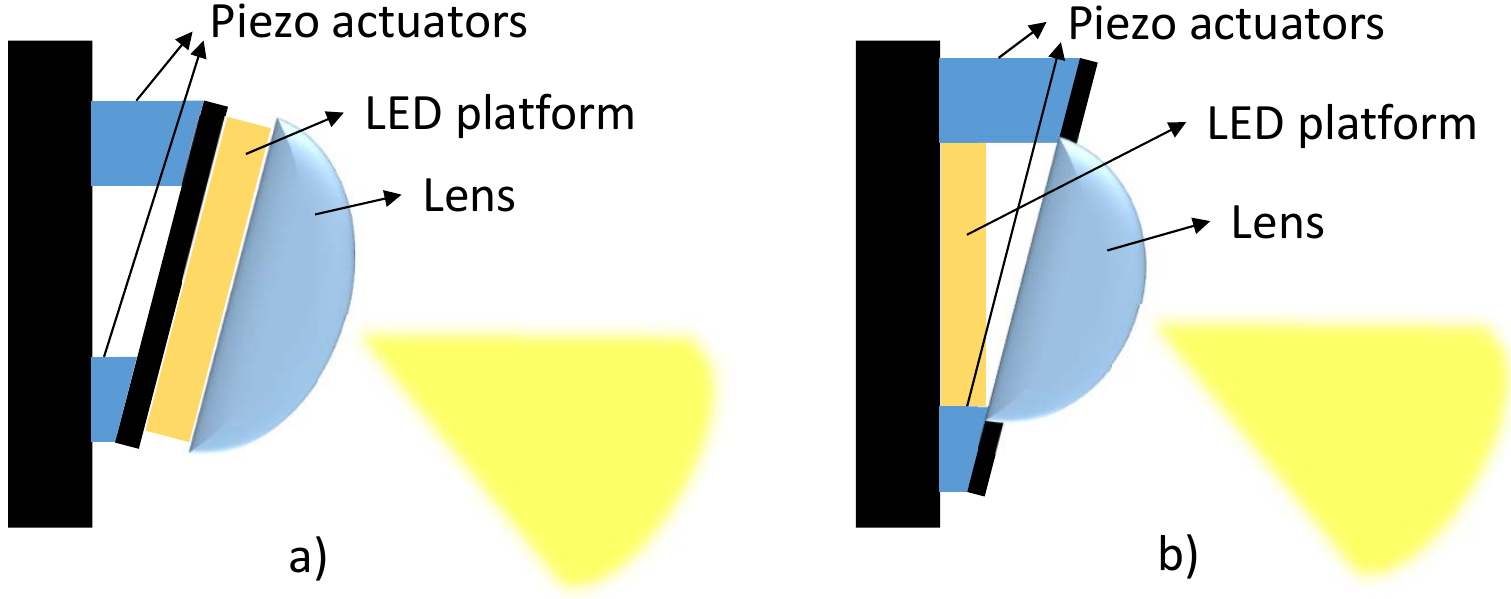}
	\caption{VLC beam steering using piezoelectric actuators. (a) The LED and the lens are steered together. (b) Only the lens is steered.}
	\label{Piezo}
\end{figure}

Piezoelectric beam steering is proposed in \cite{eroglu_VLCS} in order to track the user, improve the signal strength, and provide smoother handover between different APs. Piezo actuators convert electrical signal into precisely controlled physical displacement. This property of piezo actuators is used to finely adjust machining tools, camera lenses, mirrors, or other equipment \cite{piezoactuators}. Piezoelectric actuators can also be used to tilt LEDs or lenses to steer the beam direction towards user location. In Fig.~\ref{Piezo}, two different beam steering schemes using piezo actuators are illustrated. In the left figure, whole LED is tilted using a set of piezo actuators, while in the right figure, only the lens is steered. The setup in the right figure makes it possible to change the directivity of the light beam by shifting the  lens to forward or backward. In order to tilt an LED to any angle, two sets of piezo actuators can be used. While one provides steering on one direction, the other provides steering on a perpendicular direction.

Another method to steer LED light is to use micro-electro-mechanical system (MEMS) based mirrors \cite{Rahaim2017, 6525314, Morrison:15}, where direction of beam is controlled by changing the orientation of micro mirrors. In \cite{Morrison:15}, a setup with LEDs and MEMS mirrors is presented with steering angles of $\pm 40^\circ$ with a settling time under 5~ms, additionally featuring adaptable beam directivity. MEMS mirrors are also studied in the context of steering laser beams for indoor free space optical (FSO) communications~\cite{Varifocal, Oh:15, Knoernschild}. In this study, without assuming any of the mentioned beam steering methods, we consider a VLC AP with limited number of steerable beams that can be steered in a given range. Additionally, we consider both scenarios where 1) the beam directivity is fixed, or 2) it can be changed within a given range.

\subsection{Slow Beam Steering for Multiple Access}
We consider a model where the beam is steered so that multiple users can access the channel with time division multiple access (TDMA) without changing the beam orientation every time slot. There are two reasons not to consider changing beam orientation each time slot. The first one is, there will be time loss between each time slot for orientation change. The shortest reported settling time for LED beam steering is 5~ms~\cite{Morrison:15}, which is close to the whole TDMA frame length used for Wi-Fi systems. The second reason is that it is not possible to do such a switching without a flickering effect. Human eye can capture changes up to 200 Hz \cite{rajagopal2012ieee}, which means the whole TDMA frame length should be under 5 ms. In this paper we propose a solution where the beam is steered once, and no more steering is needed unless the location and orientation of the users change. If any user movement occurs, new steering parameters are computed and the beam is steered between TDMA frames.

\begin{figure}[tp]
	\centering
	\vspace{1mm}
	\subfigure[Single steerable beam.]{
		\includegraphics[width=1.29in] {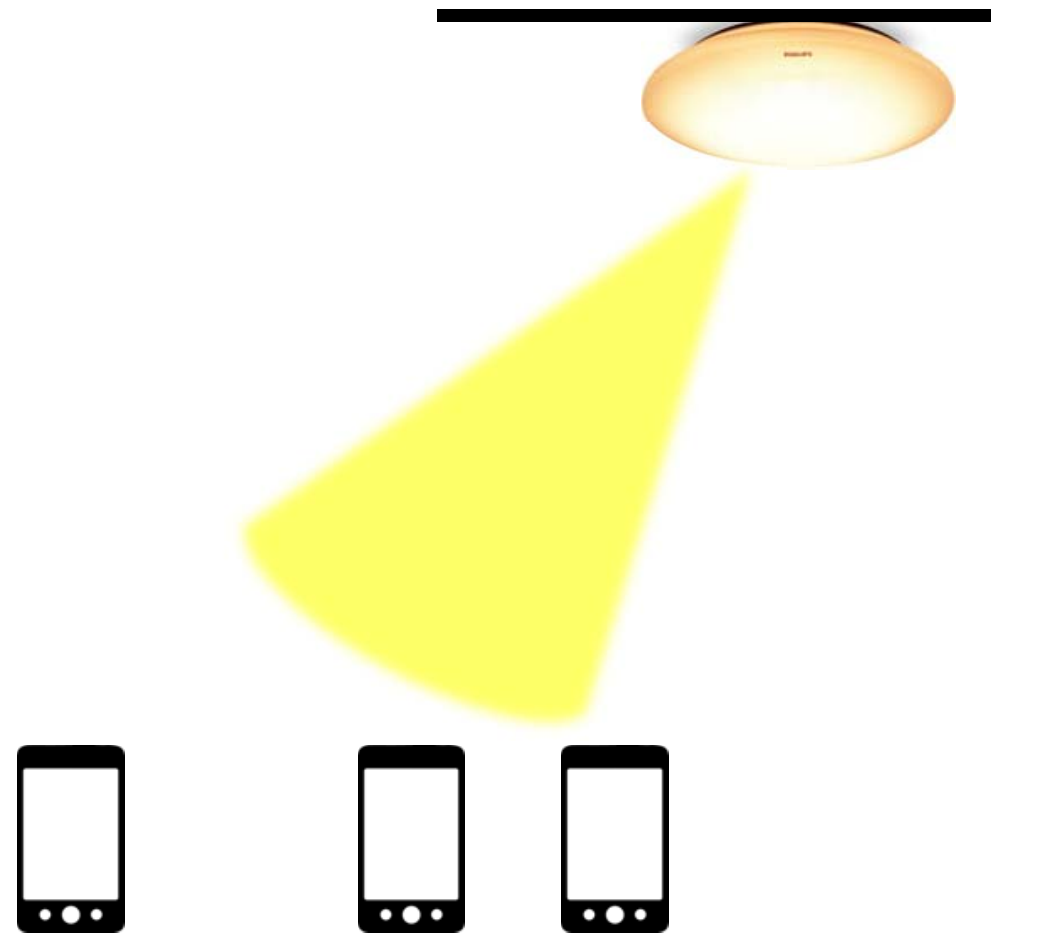}
		\label{SingleBeam}
	}
	\subfigure[Multiple steerable beams.]{
		\includegraphics[width=1.95in] {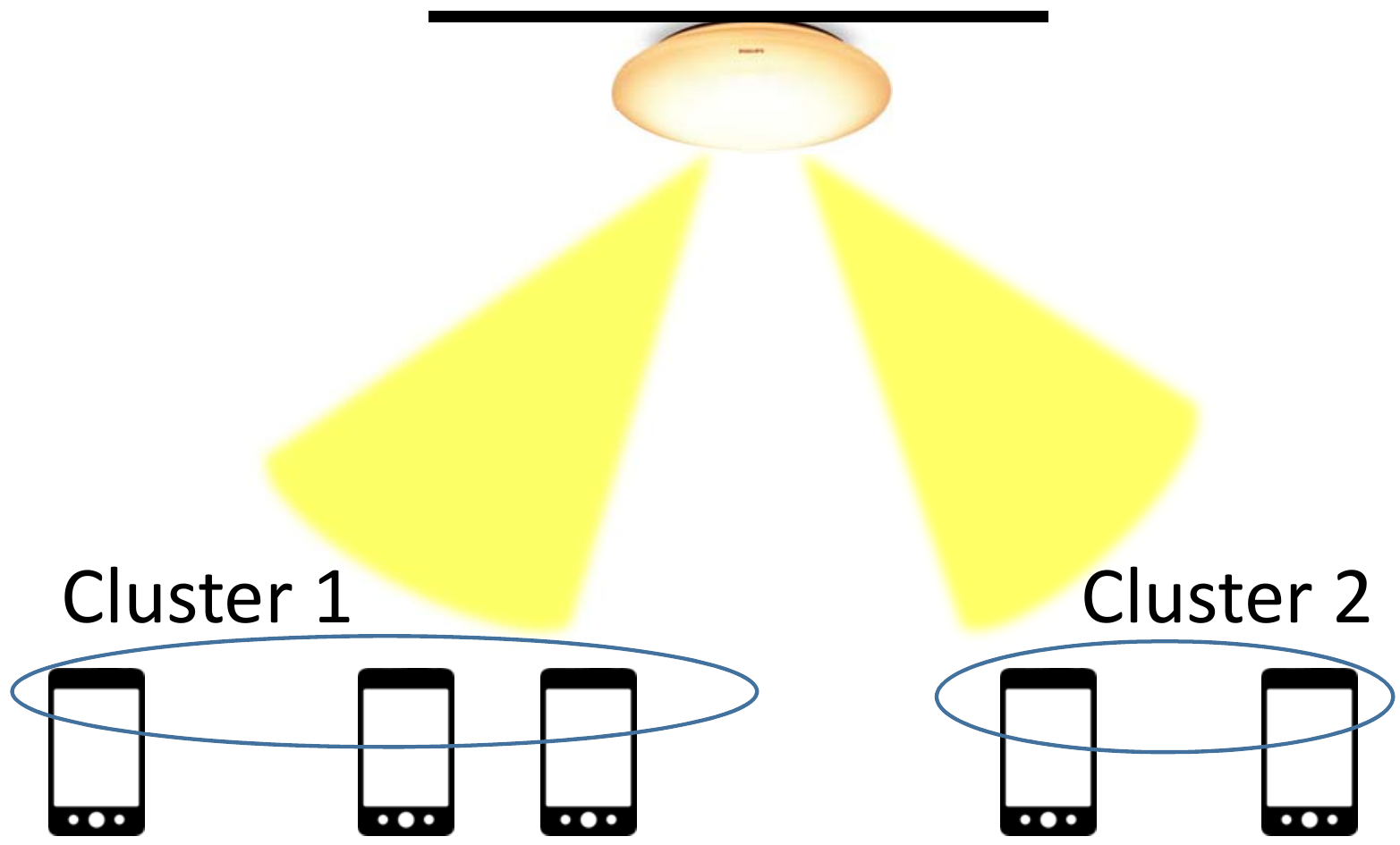}
		\label{MultiBeam}
	}
    \caption{Steering single and multiple beams to user clusters.}
	\label{SteeringScenarios}
\end{figure}

Initially, we consider an AP with a single steerable light beam and $K$ users, and Fig.~\ref{SingleBeam} shows an example scenario for $K = 3$. The AP serves all users with TDMA, and $k$th user is served with time ratio $\tau_k$. We aim at finding the steering angles and LED directivity index which maximizes logarithmic sum rate of all users. In 3D model, we need two angles to specify the orientation of the beam, which are the elevation and the azimuth angles, denoted by $\alpha$ and $\beta$, respectively. We can convert these angles to an orientation vector given as
\begin{align}
\textbf{n}_{\rm tx}&= [n_{x\rm(tx)}, n_{y\rm(tx)}, n_{z\rm(tx)}] \\ &=[\cos(\beta)\cos(\alpha),~ \sin(\beta)\cos(\alpha),~ \sin(\alpha)]. \nonumber
\end{align}
The location of the AP is $\textbf{r}_{\rm tx}=[x_{\rm tx}, y_{\rm tx}, z_{\rm tx}]$. Likewise, the location and the orientation of the $k$th user are $\textbf{r}_{k}= [x_{k}, y_{k}, z_{k}]$, and $\textbf{n}_{\rm k}= [n_{x(k)}, n_{y(k)}, n_{z(k)}]$, respectively. Then, the vector from the AP to $k$th user is  $\textbf{v}_k = \textbf{r}_{k} - \textbf{r}_{\rm tx} = [v_{x(k)}, v_{y(k)}, v_{z(k)}].$  The distance between the LED and the $k$th user is $d_k = ||\textbf{v}_k||_2$. The angle between the LED orientation and $\textbf{v}_k$ is denoted as $\phi_k$, and we can write
\begin{align}
\cos(\phi_k) = \frac{\textbf{n}_{\rm tx}^T(\textbf{r}_k-\textbf{r}_{\rm tx})}{d_k} = \frac{\textbf{v}_k^T \textbf{n}_{\rm tx}}{||\textbf{v}_k||_2} \, .
\end{align}
The angle between the receiver orientation and $\textbf{v}_k$ is $\theta_k$, and
\begin{align}
\cos(\theta_k) = \frac{\textbf{n}_k^T(\textbf{r}_{\rm tx}-\textbf{r}_k)}{d_k} = - \frac{\textbf{v}_k^T \textbf{n}_{k}}{||\textbf{v}_k||_2} \, .
\end{align}
We assume a light beam radiation follows the Lambertian pattern \cite{219552}, with $\gamma$ being the directivity index of the beam. Then, assuming the receiver has a wide field of view (FOV), we can remove the FOV constraint, and the line-of-sight (LOS) channel gain of the $k$th user can be calculated as
\begin{align}
h_k = \frac{\gamma + 1}{2\pi}& A_r\cos^\gamma(\phi_k)\cos(\theta_k)\frac{1}{d_k^2} \\
= \frac{\gamma + 1}{2\pi}& A_r \frac{ (\textbf{v}_k^T\textbf{n}_{\rm tx})^\gamma \, \textbf{v}_k^T\textbf{n}_k }{||\textbf{v}_k||_2^{n+3}} \\ 
= \frac{\gamma + 1}{2\pi}& A_r\frac{\left( v_{x(k)}n_{x(k)} + v_{y(k)}n_{y(k)} + v_{z(k)}n_{z(k)}\right)}{\left(v_{x(k)}^2 + v_{y(k)}^2 + v_{z(k)}^2\right)^\frac{\gamma+3}{2}} \label{channelGain}
\\ \times \big(v_{x(k)}\cos&(\beta)\cos(\alpha) + v_{y(k)}\sin(\beta)\cos(\alpha) + v_{z(k)}\sin(\alpha) \big)^\gamma \label{h_expanded}. \nonumber
\end{align}
Then, with a direct current (DC) biased Gaussian signal assumption \cite{7572968}, the rate of the $k$th user is given as
\begin{align}
R_k = B\log\left(1 + \frac{(rph_k)^2}{N_0B}\right),
\end{align}
where $r$ is the responsivity of photo-diode and $p$ is the transmit power of the LED, which are both considered to be constant. The $N_0$ is the spectral density of additive white Gaussian noise (AWGN), and $B$ is the communication bandwidth. Then the optimal parameters can be found solving the problem
\begin{equation}
\begin{split}
\tilde{\boldsymbol{\tau}}, \tilde{\alpha}, \tilde{\beta}, \tilde{\gamma} = {\rm arg}&\underset{ \boldsymbol{\tau}, \alpha, \beta, \gamma}{\rm ~max}~~  \displaystyle\sum_{k=1}^{K} \log(\tau_kR_{k}),\\
{\rm s.t.}~~ c_1:\quad & \alpha_{\rm min} \leq \alpha \leq \alpha_{\rm max}, \\
c_2:\quad & 0 \leq \beta \leq 360^\circ \,, \\
c_3:\quad & \gamma_{\rm min} \leq \gamma \leq \gamma_{\rm max} \,, \\ 
c_4:\quad & \sum_{k=1}^{K} \tau_k = 1 \, ,
\end{split}
\label{betaOpt}
\end{equation}
where $\boldsymbol{\tau} = [\tau_1, ... , \tau_K]$ is the time division coefficient vector which sums up to 1. To make sure all users are served and the resources are distributed fairly, the objective function is the sum of logarithmic rate instead of sum rate~\cite{3010473}. If the logarithm is removed from objective function, a single user gets all time allocation and the beam is steered towards that user, leaving other users unserved. 

\section{Proposed Solutions} \label{PropSol}
In this section, we solve the optimization problem in~\eqref{betaOpt} for a single steerable beam and multiple users, and subsequently extend the solution to the multiple beams case. 

\subsection{Solution to the Optimization Problem}
We can divide the problem in \eqref{betaOpt} to two sub-problems by separating the objective function as
\begin{equation}
\displaystyle\sum_{k=1}^{K} \log(\tau_kR_{k}) = \displaystyle\sum_{k=1}^{K} \log(\tau_k) + \displaystyle\sum_{k=1}^{K} \log(R_{k}).
\end{equation}
The first problem is $ \tilde{\boldsymbol{\tau}} = {\rm arg}~\underset{ \boldsymbol{\tau}}{\rm max}~ \log \left(\prod_{k=1}^{K} \tau_k\right),$ subject to $c_4$ in \eqref{betaOpt}. The answer to this trivial problem is $\tilde{\tau}_k = 1/K~ \forall k$. The second problem is given by
\begin{equation}
\tilde{\alpha}, \tilde{\beta}, \tilde{\gamma} = {\rm arg}\underset{ \alpha, \beta, \gamma}{\rm ~max}~~ \displaystyle\sum_{k=1}^{K} \log(R_{k}),
\label{betaOpt2}
\end{equation}
subject to $c_1$, $c_2$, and $c_3$ in \eqref{betaOpt}. The problem in \eqref{betaOpt2} is non-convex, and any gradient based optimization gets stuck in a local optima. This can be seen in the channel gain in \eqref{channelGain}, which has sine, cosine, and exponential functions of optimization parameters. In order to remove \eqref{channelGain} from the optimization problem, we follow a grid search based method and calculate the channel gain for discrete values of $\alpha, \beta$, and $\gamma$. To give an example, we separate all available range for $\alpha$ to discrete values with a small interval $\delta$ and $\boldsymbol{\alpha} = [\alpha_{\rm min}, \alpha_{\rm min} + \delta, ... , \alpha_{\rm max}]$. The sizes of $\boldsymbol{\alpha}, \boldsymbol{\beta}$ and $\boldsymbol{\gamma}$ are $s_\alpha, s_\beta$, and $s_\gamma$, respectively. We calculate the channel gain for all possible $\alpha, \beta$, and $\gamma$ combinations and form a column vector $\textbf{h}_{\boldsymbol{\alpha, \beta, \gamma}}^{(k)}$, whose length is $s_\alpha \times s_\beta \times s_\gamma$, and its indices can be mapped back to $\alpha, \beta$, and $\gamma$. Then, we can impose the optimization problem as
\begin{align}
\tilde{\textbf{d}} = {\rm arg}\underset{ \textbf{d}}{\rm ~max} \displaystyle\sum_{k=1}^{K} \log & \left(B\log\left( 1 + \frac{ \left( rp\, \textbf{d}^{T}\textbf{h}_{\boldsymbol{\alpha, \beta, \gamma}}^{(k)} \right)^2 }{N_0B}\right) \right), \nonumber \\
{\rm s.t.}~~ c_1 &:\quad \sum_{i = 1}^{s_\alpha s_\beta s_\gamma} d_{i} = 1, \label{betaOpt4} \\ \nonumber
c_2 &: \quad d_{i} = \{0,1\} ~~\forall i, 
\end{align}
where $\textbf{d}$ is a vector same size as $\textbf{h}_{\boldsymbol{\alpha, \beta, \gamma}}^{(k)}$. The constraints enforce that only one element of $\textbf{d}$ is equal to one, and the others are all equal to zero. The vector multiplication results in choosing an element of $\textbf{h}_{\boldsymbol{\alpha, \beta, \gamma}}$. The problem with \eqref{betaOpt4} is the combinatorial nature of the problem due to the integer constraint $d_{i}$'s. In order to remove the integer constraint, we modify the problem further as
\begin{align}
\tilde{\textbf{d}} = {\rm arg}\underset{ \textbf{d}}{\rm ~max} \displaystyle\sum_{k=1}^{K} \log & \left(B\log\left( 1 + \frac{ \left( rp\, \textbf{d}^{T}\textbf{h}_{\boldsymbol{\alpha, \beta, \gamma}}^{(k)} \right)^2 }{N_0B}\right) \right) \nonumber \\[2pt] 
&\qquad - \lambda ||\textbf{d}||_{0}, \\[2pt] 
{\rm s.t.}~~ &c_1:\quad \sum_{i = 1}^{s_\alpha s_\beta s_\gamma} d_{i} = 1;\quad  d_{i} \geq 0  \nonumber
\end{align}
where $||.||_0$ is the $\ell_0$ norm and $\lambda$ is a positive penalty parameter. Note that the solution set of the optimization problems (11) and (12) are the same. Thus, imposing $\ell_0$ penalty still preserves the meaning of the problem, however (12) is still combinatorial due to $\ell_0$ norm. This can be relaxed by replacing $\ell_0$ norm with a concave function (e.g. $\ell_q$ norm with $0<q<1$). Upon relaxation, the problem turns out to be non-convex and can be optimized using the majorization-minimization (MM) procedure \cite{MM_Tutorial} to a near-optimal solution. The basic idea of the MM procedure is to keep the convex part as it is and linearize the concave part of the function around a solution obtained in the previous iteration. The relaxed optimization problem with linearized $\ell_q$ norm is as follows:
\begin{align}\label{betaOpt6}
\tilde{\textbf{d}} = {\rm arg}\underset{ \textbf{d}}{\rm ~max} \displaystyle\sum_{k=1}^{K} \log & \left(B\log\left( 1 + \frac{ \left( rp\, \textbf{d}^{T}\textbf{h}_{\boldsymbol{\alpha, \beta, \gamma}}^{(k)} \right)^2 }{N_0B}\right) \right) \nonumber \\&- \lambda \sum_{i=1}^{s_\alpha s_\beta s_\gamma}W_i(t) d_i, \\ \nonumber
{\rm s.t.}~~ c_1 &:\quad \sum_{i = 1}^{s_\alpha s_\beta s_\gamma} d_{i} = 1;\quad  d_{i} \geq 0,
\end{align}
where $W_i(t) = q (d_i+\epsilon)^{q-1}$ is the weight update of the majorizer function at iteration $t$, and $\epsilon$ is a small non-negative number added to overcome the singularity issue; without $\epsilon$, $W_i(t)$ becomes undefined at $d_i = 0$. Interested readers may refer \cite{IRM_Cheth_Chandra} and references therein for further details of MM procedure. Solving (\ref{betaOpt6}) returns $\textbf{d}$, and the index of one in the $\textbf{d}$ can be mapped back to the final $\alpha, \beta$, and $\gamma$ values.

\subsection{Multiple Steerable Beams}
In this subsection, we consider a transmitter that can steer multiple beams independently, and therefore can track multiple users. In case number of users are higher than the number of beams, users can be separated to clusters, and each cluster can be served with a single beam as illustrated in Fig.~\ref{SteeringScenarios}(b). In order to cluster users, we introduce VLC user clustering (VUC) algorithm, which is a modified $k$-means clustering. Each cluster of users is served by a single beam, and the VUC algorithm assigns users to the clusters based on the signal strength received from each beam, and finds the steering parameters for each beam. The algorithm is explained in detail as follows. We assume there are $N$ steerable beams, and the steering angles and the directivity index of $n$th beam are $\alpha^{(n)}$, $\beta^{(n)}$, and $\gamma^{(n)}$, respectively. 

\begin{figure*}[tp]
	\centering
	\vspace{1mm}
	\subfigure[The AP has a single steerable beam AP.]{
		\includegraphics[width=2.07in] {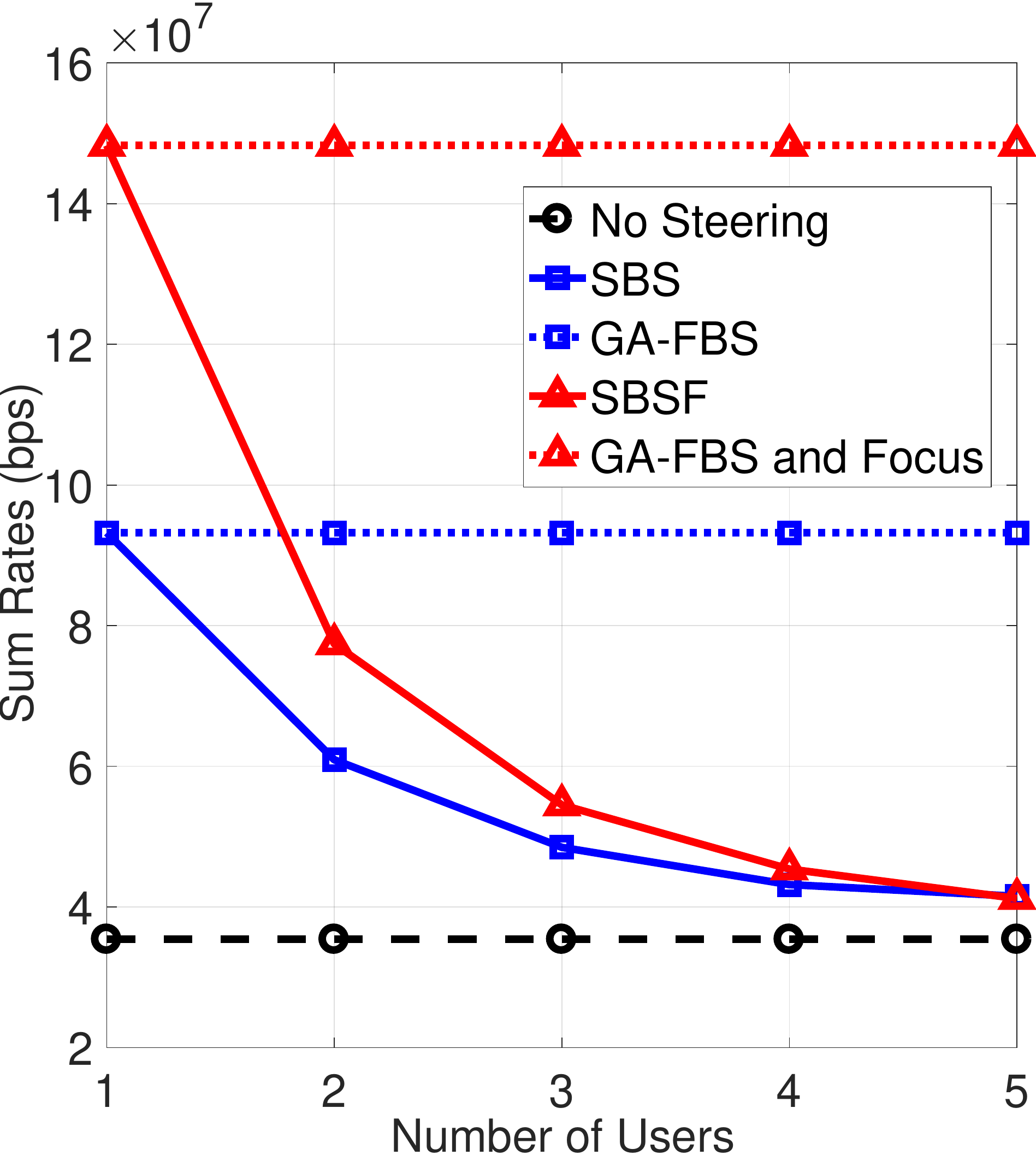}
		\label{RateSingleBeam}
	}
	\subfigure[The AP has three independently steerable beams.]{
		\includegraphics[width=2.33in] {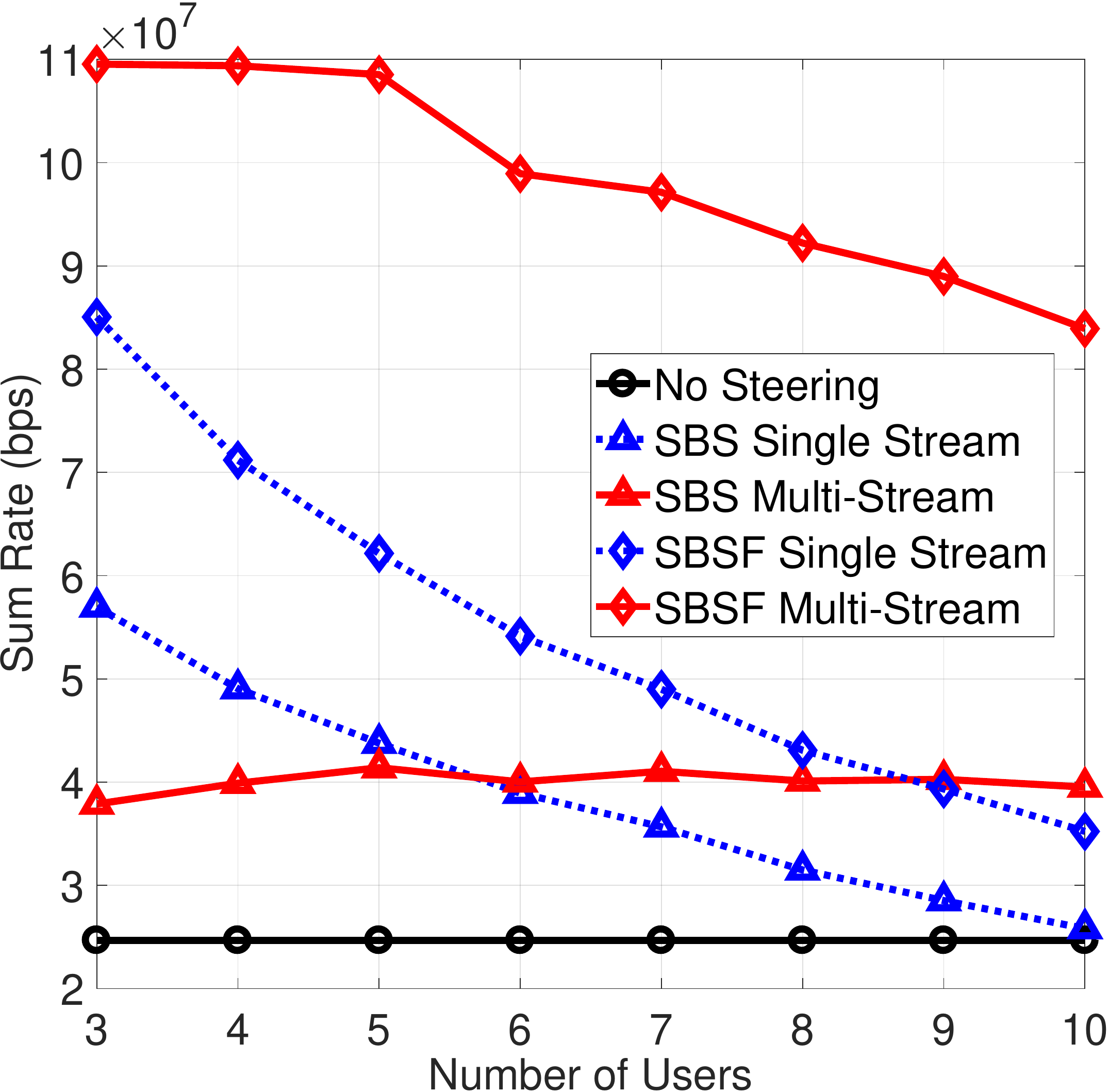}
		\label{RateClustering}
	}
    \subfigure[The AP has varying number of steerable beams serving 10 users.]{
		\includegraphics[width=2.33in] {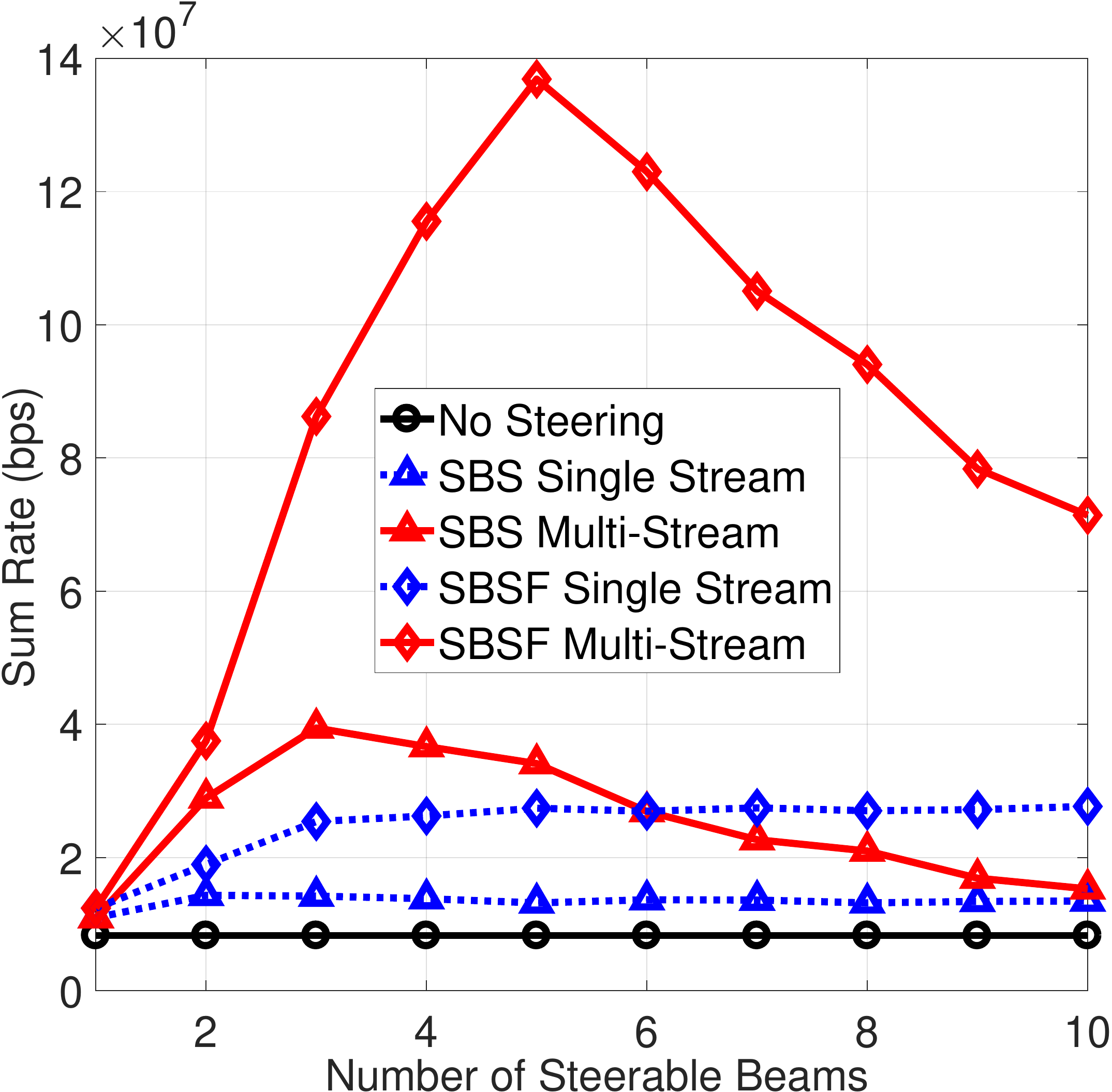}
		\label{VaryingBeams}
	}
    \vspace{-1mm}
    \caption{The sum rate of users for single and multiple steerable beams using SBS and SBSF.}
\end{figure*}

To initiate the algorithm, we randomly assign a single user to each cluster (i.e. assign first $N$ users to one cluster each). Initially there are some unassigned users, but all users will be assigned to a cluster after the algorithm is completed. We have a total of $N$ clusters, and we repeat the following steps iteratively to find conclusive clusters and cluster centers. In the first step, we calculate the steering parameters for $n$th beam, which is $\alpha^{(n)}$, $\beta^{(n)}$, and $\gamma^{(n)}$, solving the optimization problem in \eqref{betaOpt} as described in Section~\ref{PropSol}.A, for the users in $n$th cluster. We repeat it for each beam. In the second step, we assign each user to the cluster whose beam provides the maximum signal strength to the user. We repeat these two steps until the steering parameters stay the same for two consecutive iterations. 

The clustering process is summarized in Algorithm~\ref{alg}, where $J(n)$ represent the set of users assigned to $n$th cluster, and $h_{k,n}$ denotes the channel gain between $n$th  beam and $k$th user. A possible problem with this algorithm is that, a cluster may lose all its users while the algorithm is running. While this is theoretically possible, we did not encounter it in our extensive simulations with up to 10 users and 10 beams. A possible solution to this problem would be to delete the cluster and leave that LED idle, or assume a random orientation for that LED and recalculate the clusters. Restarting algorithm by assigning different initial users to clusters is another solution, since the final clusters depend on the initial cluster centers.

\begin{algorithm}[tb]
	\caption{The proposed VUC algorithm.}
	\begin{algorithmic}[1]
		\STATE Initialize: Assign user $n \rightarrow J(n) $ for $n = 1,...,N$
        \REPEAT 
        \FOR{$n$ = 1 to $N$} 
        \STATE {Solve \eqref{betaOpt} for the $n$th beam and users in $J(n)$ to find the steering parameters of $n$th beam ($\alpha^{(n)}$, $\beta^{(n)}$, and $\gamma^{(n)}$).} 
        \ENDFOR
        \FOR{$k$ = 1 to $K$} 
        \STATE { Find $n$ maximizing $h_{k,n}$, then assign user $k \rightarrow J(n)$.} 
        \ENDFOR
        \UNTIL{Steering parameters stay the same for two consecutive iterations.}
	\end{algorithmic}
	\label{alg}
\end{algorithm}

% \begin{figure}[tb]
% 	\centering
% 	\includegraphics[width = 3 in]{fig/RateSingleBeam.pdf}
% 	\caption{The sum rate of users with a single steerable beam AP.}
% 	\label{RateSingleBeam}
% \end{figure}

% \begin{figure}[tb]
% 	\centering
% 	\includegraphics[width = 3 in]{fig/RateClustering.pdf}
% 	\caption{The sum rate of users where the AP has three independently steerable beams.}
% 	\label{RateClustering}
% \end{figure}

\section{Simulation Results} \label{SimRes}
We conduct computer simulations using MATLAB, where we consider a square room of size 8~m $\times$ 8~m $\times$ 4~m. The transmitter is located in the center at ceiling level, and receivers are distributed at uniformly random locations at 0.85~m height, facing upwards. The simulation parameters are as follows: The transmit power $p$ is 1~W, the receiver responsivity $r$ is 1~A/W, the modulation bandwidth $B$ is 20~MHz, the AWGN spectral density $N_0$ is 2.5$\times 10^{-20}$ A$^2$/Hz, and the receiver surface area $A_{\rm r}$ is 1~cm$^2$. Optimization parameter limits are $\alpha_{\rm min}$ = 200$^\circ$, $\alpha_{\rm max}$ = 340$^\circ$, $\gamma_{\rm min} = 1$, and $\gamma_{\rm max}$ = 15, {and the $q$ value used for $\ell_q$ norm is 0.1}.

In Fig.~\ref{RateSingleBeam}, the sum rate of users are shown when there is only a single steerable beam. We simulate three different scenarios. The first one is labeled as ``No Steering", where the beam is not steered and faced downwards with a default directivity index $\gamma = 5$. The second one is labeled as slow beam steering (SBS), where the beam is steered as described in Section~\ref{SystemModel}.A. In this scheme we assume the directivity index cannot be changed, and equal to the default value. The third scenario is labeled as slow beam steering and focus (SBSF), where both beam orientation and directivity index are optimized. For comparison, we also consider a \emph{genie-aided} fast beam steering (GA-FBS) approach as an upper bound on the sum rate. In particular, while settling time for steering may be on the order of 5 ms in practice~\cite{Morrison:15}, we assume that we can instantaneously steer beams to each scheduled user \emph{within} an individual TDMA frame. In this scheme, the LED is completely steered towards a user for the time slot allocated to that user, and we assume steering happens with no time loss. 

Results in Fig.~\ref{RateSingleBeam} shows that when there is a single user, a significant gain on the sum rate can be achieved with steering and focusing. In this case steering angles point to the direction of the user, and the directivity index is high, since the user is on the exact direction of the beam. When the number of users increases, the total rate achievable with steering decreases. The optimization maximizes the sum of logarithm of rates to serve all users simultaneously, therefore the beam orientation does not point to a single user. Since users are not on the exact direction of the beam, the channel gains of the users decrease as the number of users increases. The sum rate for GA-FBS schemes do not decrease, because the LED is steered towards the receiving user at each time interval, and we consider the average rate over large number of user locations. 

In Fig.~\ref{RateClustering}, the sum rate of users are shown when the AP has three independently steerable beams. The transmit power of these beams are $p/3$ (versus $p$ that was used in Fig.~\ref{RateSingleBeam}) for a fair comparison. For this simulation, we consider two different multiple access schemes. The first one is labeled as single stream and shown with dashed blue lines, where all beams transmit the same signal to avoid any interference. In this scheme, the signal strength is higher, and the interference is zero. However, all the users are served with time division of a single stream, therefore they are allocated lower amount of time. In the multi-stream scheme shown with solid red lines, all beams transmit a different stream to the users assigned to them. Since Fig.~\ref{RateClustering} shows results for an AP with three independently steerable beams, multi-stream scheme has three different streams. Users assigned to the same beam are served with time division if the beam is assigned more than a single user. The sum rate decreases especially for more than three users because each user is not assigned a dedicated beam. Due to the use of spatial diversity and higher time allocation to the users, this scheme may offer higher rates than the single stream scheme.

\begin{figure}[tb]
	\centering
	\includegraphics[width = 3.1 in]{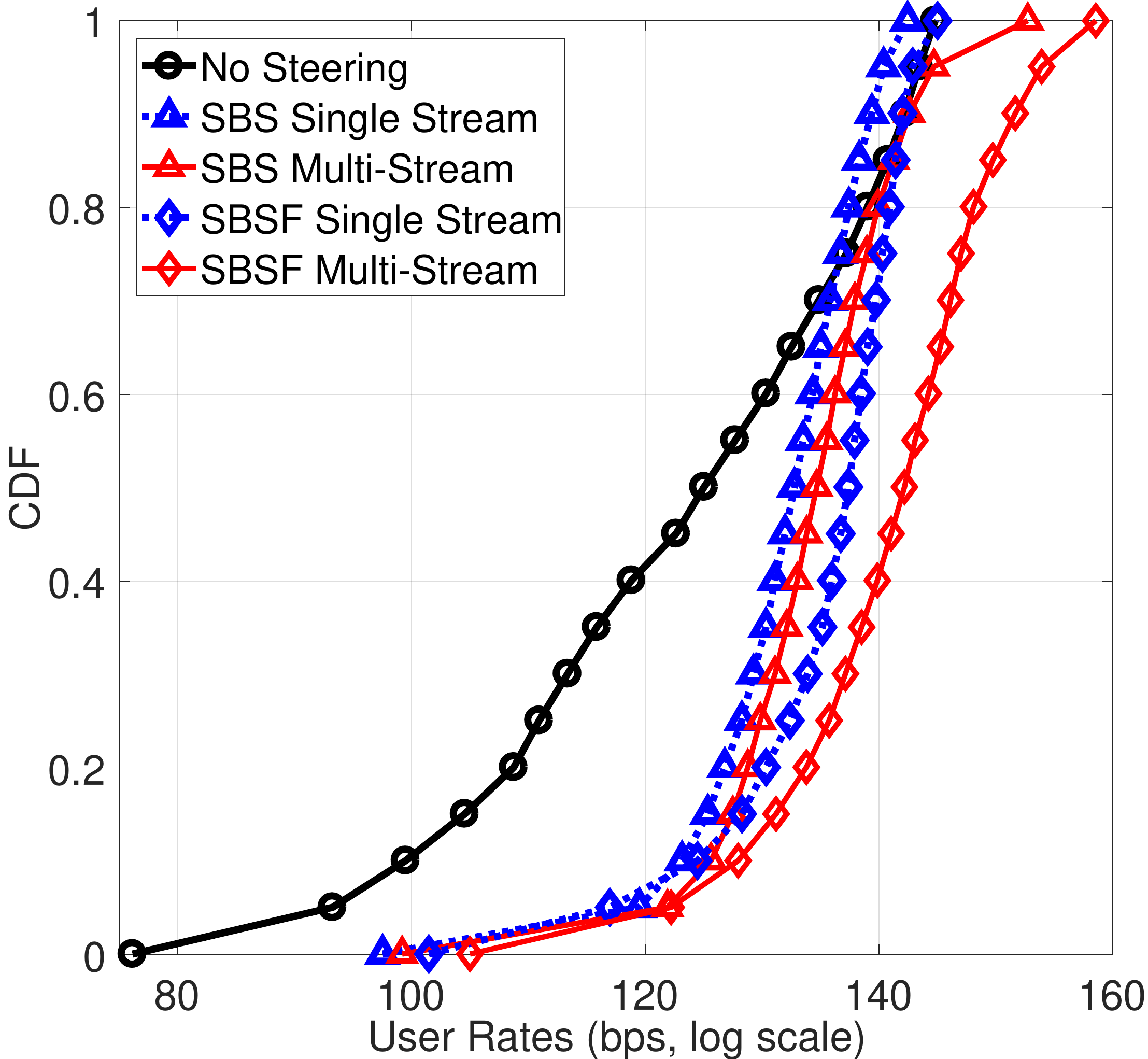}
	\caption{The CDF of individual user rates with three steerable beams and six users. The rates are converted to dB scale ($20\log_{10}(R_k)$).}
	\label{CDF}
	\vspace{-4mm}
\end{figure}

As seen in Fig.~\ref{RateClustering}, the multi-stream beams with focusing ability provide the highest sum rates. The multi-stream beams with no focusing ability do not perform well, especially with lower number of users. In this scheme a beam can cause heavy interference to other users because its directivity cannot be adjusted as needed. The slight increase in the sum rate with increasing number of users can be explained by preventing interference by clustering closer users together. With single stream scheme, the sum rate decreases and approaches to no steering scheme with increasing number of users. Since the ratio of users to the number of beams increases a lot, steering becomes less effective. Note that in Fig.~\ref{RateClustering} the sum rates do not decrease rapidly as in Fig.~\ref{RateSingleBeam}, especially sum rates of multi-stream schemes. This is due to VUC algorithm clustering users together that can receive high signal strength through a single beam. In Fig.~\ref{CDF}, the cumulative distribution function (CDF) of user rates are shown for six users and three steerable beams, as in Fig.~\ref{RateClustering}. The steering provides more uniform distribution of user rates in comparison to no steering scheme, since the optimization problem maximizes the sum of logarithm of rates and provides a fairer resource allocation.

In Fig.~\ref{VaryingBeams}, the sum rates are shown for 10 users with varying number of independently steerable beams. The transmit power of each beam is $p/N$, where $N$ is the number of beams. SBSF with multi-stream provides the highest sum rate, which is maximized at 5 beams when there are two users per beam on the average. Higher number of beams means better steering accuracy and higher received signal strength, however it also causes higher interference in multi-stream scheme and lower transmit power per beam. The ideal user count per beam ratio may change based on the size of the room or the total number of users in the room.

\section{Conclusion} \label{Conc}
In this paper we study the optimal beam steering parameters for VLC when there are higher number of users than the steerable components. We find the near-optimal steering angles and LED directivity for a single LED and multiple users. The results show that steering VLC beams and changing the directivity can improve the user rates significantly. Although serving a single user maximizes the user rates, multiple users can also be served using a single steerable beam with a significant sum rate gain over no steering scheme. In case of a multiple steerable beam setting, we cluster users and serve each cluster with a separate beam. This setting allows higher data rates by clustering close users together and providing more accurate steering. Future work includes the transmit power optimization of multiple steerable beams for maximizing the sum rate with a total power limit.

\bibliographystyle{IEEEtran} 
\bibliography{new}

\end{document}